\documentclass[cmp]{mysvjour}
\def\E{\mathbb E}

\def\N{\mathbb N}
\def\P{\mathbb P}
\def\R{\mathbb R}

\def\Z{\mathbb Z}

\def\ffi{{\varphi}}

\def\uu{{\mathbf u}}
\def\uv{{\mathbf v}}
\def\uw{{\mathbf w}}
\def\ux{{\mathbf x}}
\def\uy{\mathbf{ y}}

\def\uH{{\mathbf H}}
\def\uT{{\mathbf T}}
\def\uR{{\mathbf R}}
\def\uDelta{{\mathbf \Delta}}
\def\uU{{\mathbf U}}
\def\uV{{\mathbf V}}

\def\rV{\rm V}

\def\Bmf{\mathfrak B}

\def\bfV{\mathbf V}

\def\BSig{\rm{\mathbf \Sigma}}

\def\cA{{\mathcal A}}
\def\cB{{\mathcal B}}

\def\cE{{\mathcal E}}

\def\cM{{\mathcal M}}

\def\cS{{\mathcal S}}

\def\cR{{\mathcal R}}

\def\cZ{{\mathcal Z}}

\def\dist{{\rm{dist}}}

\def\one{{\mathbf 1}}
\def\u0{{\mathbf 0}}

\def\uf{{\mathbf f}}
\def\uu{{\mathbf u}}
\def\uv{{\mathbf v}}
\def\uw{{\mathbf w}}
\def\ux{{\mathbf x}}
\def\uy{{\mathbf y}}

\def\uA{{\mathbf A}}
\def\uB{{\mathbf B}}
\def\uC{{\mathbf C}}

\def\uM{{\mathbf M}}
\def\uH{{\mathbf H}}
\def\uG{{\mathbf G}}
\def\uI{{\mathbf I}}
\def\uS{{\mathbf S}}

\def\rx{{{\rm x}}}
\def\ry{{{\rm y}}}
\def\trho{\widetilde{\rho}}

\def\eps{{\epsilon}}
\def\om{{\omega}}
\def\Lam{{\Lambda}}
\def\Om{{\Omega}}
\def\om{{\omega}}

\def\BPsi{{\mathbf \Psi}}

\def\BLam{{\mathbf \Lam}}
\def\BLamout{{\mathbf \Lam}^{out}}

\def\TBLam{{\widetilde{\BLam}}}
\def\uTB{{\widetilde{\uB}}}
\def\utu{{\widetilde{\uu}}}
\def\tL{{\widetilde{L}}}

\def\tq{\widetilde{q}}

\def\Lamout{\Lam^{out}}

\def\BLamint{\BLam^{int}}
\def\BLamout{\BLam^{out}}
\def\TBLamout{{\TBLam}^{out}}

\def\Bout{\uB^{out}}
\def\TBout{\widetilde{\uB}^{out}}

\def\chiint{\chi^{int}}
\def\chiout{\chi^{out}}

\def\gam{{\gamma}}
\def\lam{{\lambda}}
\def\half{\frac{1}{2}}

\def\pt{\partial}
\def\pn{\par\noindent}
\def\pmn{\par\medskip\noindent}
\def\psn{\par\smallskip\noindent}

\def\myset#1{{\left\{\,#1\,\right\}}}
\def\esm#1{{\E\left[ \, #1 \, \right]}}

\def\myexp#1{{\exp\left\{\,#1\,\right\}}}
\def\pr#1{{  \P\left\{ \, #1 \, \right\}  }}

\def\dist{{\,{\rm dist}}}

\def\truc#1#2#3{\smash{\mathop{\,\, #1 \,\, }\limits^{#2}_{#3}}}

\def\tto#1{\smash{\mathop{\,\,\,\, \longrightarrow \,\,\,\, }\limits_{#1}}}

\def\mymax#1{{ \truc{\max} {} {#1}}}
\def\mysum#1{{ \truc{\sum} {} {#1}}}

\usepackage{graphics}

\usepackage{latexsym}
\usepackage{amsfonts, amssymb}
\usepackage{amsmath}
\usepackage{graphicx}
\usepackage{array}
\usepackage{rotating}
\usepackage{mflogo}

\spnewtheorem{Thm}{Theorem}[section]{\bf}{\it}
\spnewtheorem{Lem}{Lemma}[section]{\bf}{\it}
\spnewtheorem{Def}{Definition}[section]{\bf}{\it}
\spnewtheorem{Cor}{Corollary}[section]{\bf}{\it}

\numberwithin{equation}{section}

\begin{document}

\title{Anderson localization in a two-particle continuous model
with an alloy-type external potential }


\author{A. Boutet de Monvel$^1$ \and V. Chulaevsky$^2$
\and P. Stollmann$^3$\and Y. Suhov$^4$}
%
\institute{
{\scriptsize{
$^1$Institut de Math\'{e}matiques de Jussieu\\
Universit\'{e} Paris 7\\
175 rue du Chevaleret, 75013 Paris, France\\
E-mail: aboutet@math.jussieu.fr
\pmn
$^2$D\'{e}partement de Math\'{e}matiques \\
Universit\'{e} de Reims, Moulin de la Housse, B.P. 1039, \\
51687 Reims Cedex 2, France \\
E-mail: victor.tchoulaevski@univ-reims.fr
\pmn
$^3$ Fakult\"{a}t f\"{u}r Mathematik\\
Technische Universit\"{a}t Chemnitz\\
09107 Chemnitz, Germany\\
E-mail: peter.stollmann@mathematik.tu-chemnitz.de
\pmn
$^4$ Statistical Laboratory, DPMMS\\
University of Cambridge, Wilberforce Road, \\
Cambidge CB3 0WB, UK\\
E-mail: Y.M.Suhov@statslab.cam.ac.uk
}}
}

%
%

\maketitle
\begin{abstract} We establish  exponential localization for a two-particle Anderson model in a Euclidean space $\R^{d}$, $d\ge 1$, in presence of a non-trivial short-range interaction and a random external potential of the alloy type. Specifically, we prove that all eigenfunctions with eigenvalues near the lower edge of the spectrum decay exponentially in $L^2$-norm.
\end{abstract}

\section{Introduction. The two-particle Hamiltonian
in the continuum}\label{intro}
\pmn

{\bf 1A. The model.} This paper is concerned with
a two-particle Anderson model in $\R^d$ with
interaction. The Hamiltonian $\uH\left(=\uH(\omega )\right)$ is a random Schr\"{o}dinger operator
of the form
$$
\uH = -\frac{1}{2}\uDelta+\uU(\ux)+\uV(\omega;\ux)
\eqno(1.1)
$$
acting in $L_2(\R^d\times\R^d)$. This means that we consider
a pair of quantum particles, each
living in $\R^d$, in the following fashion: the joint position vector is
$\ux=(x_1, x_2)\in\R^d\times\R^d$, where each component
$x_j=\big({\rx}_j^{(1)},
\ldots,{\rx}_j^{(d)}\big)\in\R^d$ represents the coordinates of the $j$'s
particle, $j=1,2$. Next, $-\displaystyle{\frac{1}{2}}\uDelta$
is the standard kinetic energy operator obtained by adding
up the kinetic energies
$-\displaystyle{\frac{1}{2}}\Delta_j$ of individual particles and assuming
that the particles are of
identical masses. In the case of different masses,
$-\displaystyle{\frac{1}{2}}\uDelta$
would have to be
replaced by the sum $-\displaystyle{\frac{1}{2}}\sum\limits_{j=1, 2}\frac{1}{m_j}\Delta_j$,
without changing any of the analysis involved. Throughout the paper, $\Delta_j$ stands for
the Laplacian $\sum\limits_{i=1}^d\displaystyle{\frac{\partial^2}{\partial{{\rx}_j^{(i)}}^2}}$.
The interaction energy operator is denoted by $\uU(\ux )$: it is, as usually, the
operator of multiplication by a function $U(\ux )$, the inter-particle potential (which can also
incorporate a deterministic external potential). Finally, the term $\uV (\omega ;\ux )$
represents the operator of multiplication by a function
$$\ux\mapsto V(x_1;\omega )+V(x_2,\omega ),\;\;\ux=(x_1,x_2),
\eqno(1.2)
$$
where $x\in\R^d\mapsto V(x;\omega )$, $x\in\R^d$, is the random external field potential.

Our goal here is two-fold.

\psn
$\bullet\quad$\textbf{First}, we show, in a fairly general context, that the continuous version of the Multi-Scale Analysis (MSA) can be reduced, in a certain way, to its discrete counterpart, for an auxiliary lattice problem. The MSA is known to be a powerful and versatile method successfully applied to a number of spectral problems in random media. It was originally developed for lattice models (cf. \cite{FS83,FMSS85}, \cite{DK89}), and later adapted to spectral problems in Euclidean space. While the first mathematically rigorous treatment of localization in the continuum, \cite{GMP77}, deals with a different kind of models and even predates the discrete MSA story, for the continuum MSA we mention \cite{BCH97,CH94,DS01,HM84,KSS98A,KSS98B,K95,St01}, where the latter monograph contains a more detailed discussion of the literature up to the year 2000. Later developments are \cite{GK01}, \cite{BK05}  (solving the notorious problem of localization for the Bernoulli-Anderson model), and \cite{AGKW08}, where the MSA was extended to a large class of singular distributions.  It is known, however, that the latter adaptation is technically more involved than the original lattice version, thus amounting to greater complexity of the analysis of localization phenomena in continuous random environments. The reduction to an auxiliary lattice problem described in this paper "encapsulates" the so-called Geometric Resolvent Inequality (GRI) in a fairly general statement and allows a direct application of lattice techniques \textit{and some results}, technically much less involved, in a ready-made form.
In addition, we further simplify an important ingredient of the lattice MSA, following the strategy of a recent manuscript \cite{C08}. The final result is a relatively simple and short way to a proof of Anderson localization for both discrete and continuous models.

\psn
$\bullet\quad$\textbf{Second}, we combine the above mentioned techniques with those of a recent paper \cite{CS09A}, where a two-particle Anderson localization was proved for a lattice Anderson model, with Wegner-type bounds obtained in our work \cite{BCSS08} for alloy-type potentials, and thus obtain a proof of Anderson localization for a two-particle model in a Euclidean space $\R^d$ of an arbitrary dimension $d \ge 1$.

In a forthcoming paper we plan to treat a more general case of an $N$-particle system in $\R^d$, with $N >1$ and
$d\ge 1$, with a short-range interaction and subject to a random potential either of alloy-type or generated by a random field with a continuous argument. In particular, we can treat a large class of Gaussian potentials, as described in our manuscript \cite{BCS08}.
\pmn
\textit{\textbf{The plan of this paper} is as follows.
\psn
$\triangleright\;$ In \textbf{this section}, we describe our assumptions on the potential of the two-particle model, including the interaction potential and the external random potential field.
\psn
$\triangleright\;$ \textbf{In Section 2}, we discuss resolvent identities - the main technical tool of the MSA.  This is, in a sens, \textbf{the central part of the present paper}, where we describe in detail a reduction
 \footnote{Note that the results of \cite{CS09A} do not imply \textit{directly} Theorem 1.1, but, speaking informally, the main techniques are almost identical in the continuous and lattice case.}of the continuous two-particle MSA to an auxiliary lattice problem solved with the help of techniques introduced earlier in \cite{CS09A}. Indeed, a reader familiar with the latter work can easily see that the subsequent Sections 3 -- 7  follow very closely (sometimes even verbatim) respective parts of the above mentioned paper.
\psn
$\triangleright\;$ \textbf{In Section 3}, we recall the notion of "partial decoupling", or "partial separation" of two-particle cubes, introduced earlier in {\rm \cite{CS08}} in the lattice case. We also recall a useful notion of "tunneling"  in single- and two-particle boxes, which allows to conduct in a relatively simple way the inductive step of the two-particle MSA for "partially separated" boxes with no interaction.
\psn
$\triangleright\;$ \textbf{In Section 4}, the spectral localization problem is reduced to the MSA, virtually in the same way as in conventional, single-particle localization theory. We formulate here an auxiliary statement, Theorem \ref{ThmReduct}, which is a direct analog of similar statements proved in \cite{FMSS85,DK89} for single-particle tight-binding Anderson models and in \cite{CS09A} for the two-particle models (it has also been proved for $N$-particle lattice models in \cite{CS09B}). In the context of continuous Anderson models, it has been proved and used in numerous works (including the references given above; see also the monograph \cite{St01} and references therein). Speaking informally, it is neither surprising  nor novel. Nevertheless, for the reader's convenience, we give the proof of Theorem \ref{ThmReduct} in the Section \ref{SecProofThmReduct}.
\psn
$\triangleright\;$ \textbf{In sections 5, 6 and 7},  the inductive step of the two-particle MSA is made separately for three types of pairs of two-particle boxes, as described at the end of Section 4. The end of the proof of inductive estimates in Section 7 also marks the end of the proof of our main Theorem \ref{ThmMain}.
\psn
$\triangleright\;$ \textbf{In  Section \ref{SecNITRoNS}}, we prove an important auxiliary statement, which we call NITRoNS principle. This proof differs slightly from that given in \cite{CS09A}, but is fairly close to the proof of a similar statement given in \cite{CS09B}, in the context of $N$-particle systems with an arbitrary $N\ge 2$.
}
\pmn

The extension of our techniques and results to the $N$-particle models in $\R^d$, in alloy-type and more general random potentials, requires an additional argument: an induction on the number of particles, quite similar to that used in \cite{CS09B}. We prove Anderson localization for $N$-particle models in a forthcoming manuscript, in order to keep the size of this paper within reasonable limits.

Further, a reader familiar with the usual proofs of dynamical localization (cf., e.g., \cite{St01} and references therein) can see that our probabilistic estimates on finite-volume resolvents provide a sufficient input for a derivation of dynamical localization from the Multi-Scale Analysis
of 2-particle (resp., $N$-particle) interacting systems in an alloy-type random potential. This extension is also planned to be presented in a separate manuscript.

\pmn
{\bf 1B. Basic notations.} Throughout this paper, we will work with cubes in the Euclidean spaces $\R^d$, $\R^{2d}$. For our purposes, it will suffice to consider only cubes centered at lattice points $v\in\Z^d$ and, resp., $\uv\in\Z^{2d}$. With few exceptions, boldface notations correspond to "two-particle" objects, relative to $\R^{2d}$ or $\Z^{2d}$ (among these exceptions is the boldface notation for indicator functions). It is technically convenient to use the max-norm for vectors in $\R^d$ and in $\R^{2d}$: $\|x\|  = \max_{i=1, \ldots, d} |x_i|$ (idem for the norm in $\R^{2d}$). Following a tradition, we denote by $\Lam_L(u)$ (resp., $\BLam_L(\uu)$) a cube with center $u$ (resp., $\uu$) and of sidelength $2L$. In terms of the max-norm (used everywhere below), such cubes are balls of radius $L$,
with respect to the max-norm in the respective space. Further, we will also need to work with "lattice cubes" (or balls, in the max-norm) of the form
$$
B_L(u) = \Lam_L(u) \cap \Z^d, \; \uB_L(u) = \BLam_L(\uu) \cap \Z^{2d},
$$
(with "B" as in "box). Finally, we we consider "unit cells", or simply "cells", centered at lattice points:
$$
C(u) = \Lam_{1}(u) \subset \R^d, \; \uC(u) = \BLam_{1}(\uu) \subset \R^{2d},
$$
(with "C" as in "cell"). Notice that the union of all  cells $C(u), u\in\Z^d$ (resp., $\uC(\uu)$, $\uu\in\Z^{2d}$) covers the entire Euclidean space $\R^d$ (resp., $\R^{2d})$. This covering is "redundant", so that many constants are not optimal. However, in this paper we privilege the clarity of presentation to the optimality of estimates.

\pmn
{\bf 1C. Interaction and external field potentials.}
In this paper, the interaction potential is assumed
to satisfy the following property:

\textbf{(D)} \emph{Boundedness and non-negativity of $U$}:
$$
\begin{array}{cl} 0\le U(\ux) \le c, \ux\in\R^{2d}.
\end{array}
\eqno (1.3)
$$

\textbf{Remark. } Non-negativity of the interaction potential
is used to \textit{simplify} the proof of Lemma
\ref{LemNITRoNS} (see Appendix).

Further, the random external potential
$V(x;\omega )$, $x\in\R^d$, $\om\in\Omega$, is assumed to be
of alloy-type, over a cubic lattice:
$$
V(x;\om )=\sum_{s\in\Z^d}{\rV}_s(\om)\varphi_s(x-s).
\eqno (1.4)
$$
Here ${\mathcal V}=({\rV}_s,\;s\in\Z^d)$, is a family of real
random variables ${\rV}_s$ on some probability space
$(\Om,{\mathfrak B},\P )$ and $\{\varphi_s,\;s\in\Z^d\}$ is a
(nonrandom)
collection of `bump' functions $y\in\R^d\mapsto\varphi_s(y)$.
In probabilistic terms, ${\mathcal V}$ is a real-valued random
field (RF) on $\Z^d$. Physically speaking, the random variable ${\rV}_s$
represents the amplitude of an `impurity' at site $s$ of
lattice $\Z^d$ while the function $\varphi_s$ describes the `propagation' of
the impact of this impurity across $\R^d$.

To avoid excessive technicalities concerning self-adjointness of the
Hamiltonian $\uH_{\BLam}$, we impose conditions \textbf{(E1)}--\textbf{(E4)}
below:

\textbf{(E1)} \emph{Boundedness and non-negativity of ${\rV}_s$}:
$$
\sup_{s\in\Z^d}{\rV}_s =:M<\infty,
\; \inf_{s\in\Z^d}{\rV}_s \ge 0
\eqno (1.5)
$$
Again, non-negativity plays a technical role and
is not crucial for the main result. The
boundedness condition for the random variables ${\rV}_s$ can
be replaced by finiteness of some moments $\esm{|{\rV}_s|^n}$.
\pmn

\textbf{(E2)} \emph{Boundedness, non-negativity and compact support
of $\ffi_s$}: the bump functions are non-negative functions, with  bounded support, such that
$$
\truc{\sup}{}{x\in\R^d}
\left[ \sum\limits_{s\in\Z^d}\;\varphi_s(x-s)\right]<+\infty,
\;\;\forall\;x\in\R^d .
\eqno (1.6a)
$$
and $\exists$ $R\in (0,\infty )$ with
$$
\varphi_s(y)=0\;\hbox{ whenever }\;||y||_{\max}>R.
\eqno(1.6b)
$$

We will also need

\textbf{(E3)} \emph{Covering condition for $\ffi_s$}\footnote{This condition can be relaxed essentially in the same way as in the single-particle theory.}:
$$
\sum\limits_{s\in\Lam_L(u)\cap\Z^d}\;\varphi_s(x-s)
\geq\;1,\;\;\forall\;L\geq 1,\;u\in\R^d,\;x\in\Lam_L(u).
\eqno (1.7)
$$

We stress that we do not use independence of the random
variables ${\rV}_s$ for different sites $s\in\Z^d$. What
we need is a regularity property for the induced conditional
marginal distribution; see below.

Given a site $s\in\Z^d$, consider the conditional distribution function
$$F\left(\,{\ry} \big| \Bmf_s^{\rm c}\right):=
\P\big({\rV}_s<{\ry}\big|\Bmf_s^{\rm c}\big),
\eqno (1.8)
$$
relative to the sigma-algebra $\Bmf_s^{\rm c}$ generated
by the random variables $V_t,\;t\in\Z^d\setminus\{s\}$. The following condition is general enough so as to cover a
large class of external potentials, e.g., the absolute
value of a regular Gaussian random field as well as some Gibbsian random fields.
Notice, however, that it can be  relaxed  further. In this paper,
we do not seek maximal generality, preferring simplicity of
presentation.

\textbf{(E4)} \emph{Uniform marginal control of
$F\left(\,{\ry} \big| \Bmf_s^{\rm c}\right)$}: the conditional distribution function $F\left(\,{\ry} \big| \Bmf_s^{\rm c}\right)$
is H\"{o}lder-continuous: for some $b >0$ and all $\eps\in(0,1)$,
$$
\begin{array}{l}\nu (\eps )
:= {\operatornamewithlimits{\sup}\limits_{s\in\Z^d}}\;\;
{\operatornamewithlimits{\sup}\limits_{\ry\in \R}}\;\;
{\operatornamewithlimits{{\rm{sup\,ess}}}\limits_{
{\bfV}_{\{s\}^{\rm c}}}}
\;\Big[F\left(\,{\ry}+\eps  \big| \Bmf_s^{\rm c}\right)-
F\left(\,{\ry} \big| \Bmf_s^{\rm c}\right)\Big]\,\le \eps^b.\end{array}
\eqno (1.9).
$$
\pmn
\textbf{Remark. } The main results of this paper remain valid under a much weaker assumption of \textit{log}-H\"{o}lder continuity of the conditional distribution function: $\nu(\eps) \le Const \, \ln^{-A} \eps^{-1}$, with sufficiently large $A>0$. Note also that in \cite{CS09A} a mush stronger assumption was made: existence and boundedness of the \textit{marginal density} $p_V$ of the external potential $V$, supposed to have independent identically distributed lattice in the framework of the lattice (tight binding) Anderson model considered in \cite{CS09A}. The only reason why the absolute continuity of the random variables $V(x;\om)$ was supposed is that allowed to apply directly earlier results from the single-particle theory proved by Aizenman et al. and which required  the existence and boundedness of the density $p_V$. Specifically, these results were used in the proof of an analog of our Lemma \ref{LemNITRoNS}, which we call here the "NITRoNS principle". Later, the proof and even the formulation of the NITRoNS principle was simplified and generalized in \cite{CS09B}, without using results from single-particle theory. This simplification allows also to substantially relax the assumptions upon the regularity properties of the distribution of the values of the random potential $V$.

\pmn
{\bf 1D. Main result.} The main result of this paper is the
following
\pmn

\begin{Thm}\label{ThmMain} Consider  the operator $\uH$ from {\rm{(1.1)}}.
Under conditions {\rm{(D)}} and {\rm{(E1)}}--{\rm{(E4)}}, it admits a
unique self-adjoint extension from the set of $C^2$-functions with
compact support in $\R^d\times\R^d$. This self-adjoint extension,
again denoted by $\uH$, is a random positive-definite operator with the
following property. Let $E_0^*$ be the lower edge of the spectrum of the operator
$-\frac{1}{2}\uDelta+\uU(\ux)$. There exists $E_1^*>E_0^*$ such that the
spectrum of $\uH$ in $[E_0^*, E_1^*]$ is pure point with $\P$-probability one.
Moreover,
there exists a (non-random) constant $m>0$ such that
for each eigenfunction $\BPsi_j(\ux;\om)$ with eigenvalue $E_j\in [E_0^*,E_1^*)$
and $\forall$ $\uv\in\Z^d\times\Z^d$,
$$
 \| \one_{\uC(\uv) }\BPsi_j(\cdot;\om)\|_{L^2(\R^{2d})} \le C_j e^{-m\|\uv\|}.
\eqno(1.10)
$$
where $C_j=C_j(\om )\in (0,+\infty )$ is a random constant
varying with $j$.
\end{Thm}
\pmn
\textbf{Remark. } The spectrum of operator $\uH(\om)$ may have empty intersection with $[E_0^*, E_1^*]$, in which case the assertion is satisfied automatically; to exclude this case one could assume that $0$ belongs to the support of the law of every $V_s$ or, more precisely, that the conditional distribution function of each $V_s$  is strictly
monotone in some interval $[0,\delta]$, $\delta >0$. Finally, observe that the essential spectrum of operator
$\widetilde{\uH} = -\frac{1}{2}\uDelta+\uU(\ux)$ with a short-range interaction $\uU$ starts at $0$. Indeed,  there exist arbitrarily large cubes $\BLam_L(u_1,u_2) = \Lam_L(u_1)  \times \Lam_L(u_2)$ with $\Lam_L(u_1)  \times \Lam_L(u_2)=\varnothing$, on which $\widetilde{\uH} = -\frac{1}{2}\uDelta$, since $U(x_1,x_2)\Big|_{\BLam_L(u_1,u_2)}=0$. Recall also that, by virtue of an earlier result proved by  Klopp and Zenk, (cf. \cite{KZ03,KZ09})  the integrated density of states, or the limiting eigenvalue distribution function, is the same for
the (nonrandom) operator $\widetilde{\uH}$ and for the ensemble of random operators $\uH(\om)$ (under certain conditions).
\pmn

The proof of Theorem \ref{ThmMain} is based on the analysis of the operators
$\uH^{\BLam}$, the finite-volume versions of $\uH$. More precisely, let $\BLam = \BLam_L(\uu)$ and consider the operator
$\uH^\BLam$ in $L_2(\BLam )$ defined as in (1.1):
$$
\uH^{\BLam} = -\frac{1}{2}\uDelta^{\BLam}+\uU(\ux)+\uV(\omega;\ux)
\eqno(1.11)
$$
where $\uDelta^{\BLam}$ stands for the kinetic energy operator
in $L_2(\BLam )$ with Dirichlet's boundary conditions on $\partial\BLam$.
Under assumptions (D) and (E1)--(E4), there exists
a unique self-adjoint extension of $\uH^{\BLam}$ from the set
of $C^2$-functions vanishing in a neighbourhood of
the boundary $\partial\BLam$; we again denote it by $\uH^\BLam$.
Then $\uH^\BLam$ is a random positive-definite operator with
pure point spectrum which will be denoted by  $\BSig\,(\uH^\BLam )$. Furthermore,
the resolvent $(\uH^\BLam -z\uI )^{-1}$, for $z\not\in\BSig\,(\uH^\BLam )$, is a compact integral operator in $L_2(\BLam )$. The MSA is an asymptotical study of
Green's functions $\uG^\BLam (\ux ,\uy;z)$, i.e. the kernels
of operators $(\uH^\BLam -z\uI )^{-1}$ as $\BLam\nearrow\R^d\times\R^d$.

\section{Resolvent inequalities}\label{SectGRI}

As is well-known by now, the MSA consists of a certain number of probabilistic estimates, proved inductively (decay estimates of Green functions) or for all scales at once (Wegner-type bounds), combined with "deterministic", functional-analytic inequalities for resolvents in finite cubes. In this section, we discuss such resolvent inequalities and show that they can be essentially reduced to those for some auxiliary functions defined on a lattice. This does not mean that we reduce the spectral problem in question, formulated for a differential Schr\"{o}dinger operator, to that for a finite-difference operator. However, the deterministic component of the MSA scheme  proposed in this paper deals with lattice functions.

First of all, we have to define standard notions of "non-resonant" and "non-singular" cubes; these definitions clearly go back to the well-known paper by von Dreifus and Klein \cite{DK89}. Note, however, that these definitions do not depend upon a particular structure of the potential of the Schr\"{o}dinger operator. So, they apply, formally, both to single- and to multi-particle Hamiltonians. Definition \ref{DefNS} is slightly modified, as compared to many well-known papers on applications of the MSA, for the reason explained below. In Definition \ref{DefNR}, we introduce a requirement somewhat stronger than usual, but the reader familiar with the MSA knows that it is this, stronger requirement is usually made, in order to avoid "too singular" singular finite boxes. The form of our Definition \ref{DefNR} gives rise to shorter arguments.
\pmn
For the sake of brevity, we introduce below the notions of "resonant" and "singular" cubes only in dimension $2d$, i.e., in the "two-particle" context. These notions are defined in dimension $d$ (in the context of single-particle problems) in the same way, by replacing $2d$ with $d$.

\pmn
\begin{Def}\label{DefNR}
Consider a cube $\BLam_L(\uu)$ centered at a point $\uu\in\Z^{2d}$. It is called $E$-non-resonant ($E$-NR, in short) if for any $\ell \ge L^{1/\alpha}$ and any cube of sidelength  $2\ell$,
$\BLam_\ell(\uv) \subseteq \BLam_L(\uu)$, with center $\uv\in\Z^{2d}$, the following bound holds true:
$$
\dist[ E, \Sigma(\uH_{\BLam_\ell(\uu)}) ] \ge e^{-\ell^\beta}, \; \beta = 1/2,
\eqno(2.1)
$$
Otherwise, it is called $E$-resonant ($E$-R).
\end{Def}

Now consider a cube $\BLam_L(\uu)$ and set
$$
\begin{array}{l}
\BLamout = \Lam_L(u) \setminus \Lam_{L-2}(u).
\end{array}
\eqno(2.2)
$$
\begin{Def}\label{DefNS}
A cube $\BLam_L(\uv)$ is called $(E,m)$-non-singular ($(E,m)$-NS, in short) if
for any $\uw\in \Lamout \cap\Z^d$,
$$
 \|  \one_{\uC(\uv)} \, \uG_{\BLam_L(\uv)}(E) \, \one_{\uC(\uw)} \| \le e^{-\gam(m,L)},
\eqno(2.3)
$$
where
$$
\gam(m,L) := m L\left(1 + L^{-1/4}\right).
\eqno(2.4)
$$
\end{Def}
\pmn
\textbf{Remark for readers familiar with the traditional MSA.} The reason why we describe the decay of the Green function $\uG_{\BLam_L(\uv)}(E)$ in terms of $e^{-\gamma(m,L)}$ instead of  the more traditional $e^{-mL}$ is that it allows to avoid "mass rescaling" when passing from a scale $L_k$ to $L_{k+1}$. Indeed, it is straightforward that if  the positive numbers $m_k$ and $m_{k+1}$ are related by $m_{k+1} \ge m_k (1 - L_k^{-1/2})$, then
$$
\begin{array}{l}
\gamma(m_{k}, L_k) (1 - L_{k}^{-1/2}) = m_{k}(1 + L_k^{-1/4})  (1 - L_{k}^{-1/2}) \\
=  m_{k}(1 + L_k^{-1/4} - L_k^{-1/2} - L_k^{-1/8}) > m_k( 1 + L_{k+1}^{-1/4}) = \gamma(m_{k}, L_{k+1}),
\end{array}
$$
provided that $L_k$ is large enough, so that $L_k^{1/2} - 2 > L_k^{1/8}$. Therefore, having a decay exponent  $\gam(m,L_k)$ at scale $L_k$, a traditional rescaling gives a decay exponent \textit{larger} than $\gam(m,L_{K+1})$  at a larger scale $L_{k+1}$. This means precisely that we can use the decay exponent $\gam(m,L_{k+1})$ without rescaling the value of the parameter $m$ (the function $\gam(m,L)$ automatically takes care of it).
\pmn

Consider two embedded cubes, $ \BLam_L(\uu)\subset\BLam' \subset\R^{2d}$, and measurable subsets thereof, $\cA \subset \BLamint := \BLam_{L/3}(\uu)$, $\cB \subset \BLam \setminus \BLam'$. For our purposes, it suffices to consider $\cA$ and $\cB$ of cubic form. Then the well-known resolvent identity for Schr\"{o}dinger operators in $\R^{2d}$ combined with commutator estimates  implies the geometric resolvent inequality of the following form (cf. \cite{St01}):
\psn
$$
\begin{array}{ll}
\textbf{(GRI):} \qquad\qquad\qquad\qquad\qquad\qquad\qquad\qquad\qquad\qquad\qquad\qquad\qquad\qquad\qquad\\
     \qquad\qquad \|  \one_\cB \, \uG_{\BLam'}(E) \, \one_\cA \|
     \le C_{geom}  \|  \one_{\cB} \, \uG_{\BLam'}(E) \, \one_{\BLamout} \|
     \,  \|  \one_{\BLamout} \, \uG_{\BLam} \, \one_\cA \|
\end{array}
\eqno(2.5)
$$

In the above inequality and below, we always use the ${L^2(\R^{2d})}$-norms, omitting the subscript ${L^2(\R^{2d})}$ for notational brevity.

In what follows, we will always use the GRI in the context where the subsets $\cA, \cB$  appearing in the (2.5) are finite unions of unit cells $\uC(\uu)$ introduced in Subsection 1A; recall that centers of the unit cells are points of the lattice $\Z^{2d} \subset \R^{2d}$. This will allow an effective "discretization" of most important functions, including Green functions, defined in the continuous space, and reduce most of our estimates to those for functions defined on a lattice. In turn, this leads to a unified approach to Anderson localization in discrete and continuous models.

\psn
\textbf{Remark. } Our methods admit a natural extension to other $d$-dimensional lattices $\cZ \subset \R^d$, i.e. additive subgroups $\cZ$ of the group $\R^d$ generated by $d$ linearly independent vectors $\mathbf{e_1}, \ldots, \mathbf{e_d}\in\R^d$.

\subsection{Discretized integrated  Green functions}

Given a point $\utu\in\Z^{2d} \subset \R^{2d}$ and a positive integer $\tL$, consider the cube $\TBLam=\BLam_{\tL}(\utu)$ and  the lattice cube $\uTB = \uB_{\tL}(\utu)$. Further, pick a point $\uu\in \TBLam$ and a positive integer $L < \tL$ such that $\BLam_L(\uu) \subset \TBLam_{\tL-3}(\utu)$. As above, introduce annular areas in the Euclidean space $\R^{2d}$,
$$
\TBLamout = \BLam_{\tL}(\utu)\setminus \BLam_{\tL-2}(\utu),
\quad \BLamout = \BLam_{L}(\uu)\setminus \BLam_{L-2}(\uu),
$$
and in the lattice $\Z^{2d}$,
$$
\TBout =\TBLamout \cap \Z^{2d},
\quad \Bout = \BLamout \cap \Z^{2d}.
$$

Then it is clear that
$$
\TBLamout \subset \bigcup_{\uv\in \TBout}  \uC(v),
\quad \BLamout \subset \bigcup_{\uv\in \Bout}  \uC(v),
$$
so that for the indicator functions we obtain
$$
\one_{\TBLamout} \le  \sum_{\uv\in \TBout}  \one_{\uC(\uv)},
\quad \one_{\BLamout} \le  \sum_{\uv\in \Bout}  \one_{\uC(\uv)}.
$$
Therefore, (GRI) implies that for any $\uw\in \TBLamout \cap \Z^{2d}$, we have the following inequality:
$$
\begin{array}{l}
\|  \one_{\uC(\uu)} \, \uG_{\TBLam}(E) \, \one_{\uC(\uw)} \| \\ \\
     \le  C_{geom}  \mysum{\uv \in \Bout}
     \|  \one_{\uC(\uu)} \, \uG_{\BLam}(E) \, \one_{\uC(\uv)} \|
     \,  \|  \one_{\uC(\uv)} \, \uG_{\TBLam} \, \one_{\uC(\uw)} \|,
\end{array}
$$
Given any pair of lattice points $\uu,\uv\in\Z^{2d}$ and $\uTB = \TBLam\cap\Z^{2d}$, denote
$$
\cR_{\uTB}(\uu,\uv; E) = \cR_{\TBLam}(\uu,\uv; E) :=  \|  \one_{\uC(\uu)} \, \uG_{\TBLam} \, \one_{\uC(\uv)} \|.
$$
With this notation, the above equation  takes the following form (which we will call \textit{"Lattice Geometric Resolvent Inequality"} (LGRI, for short), in order to distinguish   it from the GRI in Euclidean space:
$$
\begin{array}{ll}
\textbf{(LGRI):} \hfill\\
     \qquad\qquad\;\;\;  \cR_{\TBLam}(\uu,\uw;E) \le C_{geom} \mysum{\uu \in \BLamout \cap \Z^{2d} }
      \cR_{\BLam}(\uu, \uv; E) \, \cR_{\TBLam}(\uv, \uw; E)
\end{array}
\eqno(2.6)
$$
or, equivalently,
$$
\begin{array}{ll}
     \qquad\qquad\;\;\;  \cR_{\uTB}(\uu,\uw;E) \le C_{geom} \mysum{\uu \in \Bout }
      \cR_{\uB}(\uu, \uv; E) \, \cR_{\uTB}(\uv, \uw; E)
\end{array}
\eqno(2.6')
$$
\pmn
Now the analogy with the lattice version of the GRI (see, e.g., \cite{DK89}) is straightforward; the only difference is a geometrical constant in the RHS. However,  with the first factor in the terms of the sum small enough, this constant will not require a substantial modification of the lattice MSA technique.

A reader familiar with the MSA can easily see now that the central problem of the Multi-Scale Analysis for our model in the Euclidean space $\R^{2d}$ is essentially reduced to the analysis of the decay properties of the functions
$\cR_{\uB_L(\uv)}(\uv, \uw; E)$ defined on $\Z^{2d} \times \Z^{2d}$. On the other hand, the spectral problem for the operator $\uH(\om)$ is \textit{not} formally reduced to that for a tight-binding Hamiltonian in $\ell^2(\Z^{2d})$.
\pmn

It is worth mentioning that our reduction of the MSA in Euclidean space to an auxiliary lattice problem is not contingent upon a particular structure of the random external potential. The fact that the centers of the scatterers of the alloy-type potential considered in this paper form the same cubic lattice $\Z^d$ as the centers of unit cells $\uC(v)$ is a mere coincidence. Moreover, the above mentioned discretization can be used, with no modification, in the case where the random potential $V(x;\om)$ is a random field with continuous argument (e.g., a regular Gaussian field with continuous argument, as in our recent manuscript \cite{BCS08}).

Working with lattice cubes (and, more generally, lattice sets) $\BLam$, we will use traditional notations for the inner boundary $\pt^-\BLam$, exterior boundary $\pt^+\BLam$, and "full" boundary $\pt\BLam$, defined as follows:
$$
\begin{array}{l}
\pt^- \Lam = \myset{x:\in\Lam:\, \dist[x, \Z^d \setminus \Lam ] = 1}, \\
\pt^+ \Lam =\myset{x:\in \Z^d \setminus \Lam:\, \dist[x, \Lam ] = 1}, \\
\pt \Lam =\myset{(x,x'): \|x-x'\|=1, x\in\pt^-\Lam, x'\in \pt^-\Lam}.
\end{array}
$$

\subsection{\textbf{LGRI} for NS-boxes}
\label{ss:GRI_NS}

Fix a box (i.e., a lattice cube) cube $\uB_L(\uu)$ and a lattice point $\uy\in\pt^- \uB_L(\uu)$. Assume that $\uB_L(\uu)$ does not contain any $(E,m)$-singular box $\uB_\ell(\uv)$. Then the LGRI (2.6') implies for such a box that for any $\uB_\ell(\uv) \subset \uB_L(\uu)$
$$
 \cR_{\uB_L(u)}(\uu,\uy;E)| \le \tq \mymax{\uv\in \pt^+\uB_\ell(\uu)} \; \cR_{\uB_L(\uu)}(\uv,\uy;E).
\eqno(2.7)
$$
with
$$
\tq = e^{-m\ell} \, C_{geom}| \pt \Lam_\ell(u)| \, \le e^{-m\ell} C'\ell^{d-1},
$$
where $C'$ is another geometrical constant.

\subsection{\textbf{LGRI} for non-resonant singular cubes}
\label{ss:GRI_S}

Now consider a situation where a box $\uB_L(\uu)$ contains a $(E,m)$-singular box $\uB_\ell(\uv)$, but
\psn
(i) any box $\uB_\ell(\uv')$ such that $\|\uv - \uv'\|=2\ell-1$, i.e.
$\dist[ \uB_\ell(\uv), \uB_\ell(\uv') ] = 1$, is $(E,m)$-non-singular;
\psn
(ii) all boxes $\uB_s(\uw) \subset \uB_L(\uu)$ with $s \in [\ell,L]$ are $E$-non-resonant.
\psn
Fix a point $\uy\in \pt^-\BLam_L(\uu)$. Then the LGRI  implies that for any
$\uB_\ell(\uv) \subset \uB_L(\uu)$
$$
 \cR_{\uB_L(\uu)}(\uv,\uy;E) \le C_{geom} e^{\ell^\beta} |\pt^+\Lam_\ell(\uv)|  \; \mymax{\uw: \|\uw - \uv\|=2\ell-1} \; \cR_{\uB_L(\uu)}(\uw,\uy;E)
 \eqno(2.8)
$$
Applying the LGRI  to all neighboring boxes $\uB_\ell(\uw)$, we come to the following bound:
$$
 \cR_{\uB_L(\uu)}(\uv,\uy;E) \le q \, \mymax{\uw:\; \ell  \le \|\uw - \uv\|=2\ell-1} \; \cR_{\uB_L(\uu)}(\uw,\uy;E)
 \eqno(2.9)
$$
with
$$
q = e^{-m\ell} e^{\ell^\beta} C'' \ell^{d-1}.
$$
Indeed, it is plain that all above mentioned boxes $\uB_\ell(\uw)$ are contained in a "layer"
of width $2\ell-1$ around the box $\uB_\ell(\uv)$,
$$
\{\uw:\; \ell  \le \|\uw - \uv\|=2\ell-1 \}.
$$
More generally, given a positive number $A<\infty$, suppose that a box $\uB_\ell(\uv)$ is $(E,m)$-singular, but:
\psn
(a) the box $\uB_{A\ell}(\uv)$ is $E$-non-resonant;
\psn
(b) any box $\uB_\ell(\uw)$ such that $\dist[\uB_\ell(\uv), \uB_\ell(\uv')] = 1$, is $(E,m)$-non-singular.

\psn
Then the analog of (2.9) reads as follows:
$$
 \cR_{\uB_L(\uu)}(\uv,\uy;E) \le q \, \mymax{\uw: \|\uw - \uv\|=(A+1)\ell-1} \; \cR_{\uB_L(\uu)}(\uw,\uy;E)
\eqno(2.10)
$$
with
$$
q = e^{-m\ell} e^{\ell^\beta} C''' \ell^{d-1},
$$
where $C''' = O(A^{d-1})$.

Observe that $\tq \le q$, so that the above Eqn (2.7) implies a slightly weaker inequality
$$
 \cR_{\Lam_L(\uu)}(\uu,\uy;E)| \le q \mymax{\uv\in \pt^+\Lam_\ell(\uu)} \; \cR_{\Lam_L(\uu)}(\uv,\uy;E),
\eqno(2.11)
$$
\pmn
with the same value of $q$ as in (2.9), (2.10). We see that the difference between cases (2.7) and (2.10)
resides in the form (and size) of the "reference set" of points $\uw$ used in  these recurrent relations.

\subsection{Clustering of disjoint singular boxes}

Fix a box lattice box $\uB_L(\uu)$ and suppose that it contains some singular boxes of size $\ell$.  In order to be able to apply to a given singular box $\uB_\ell(\uv^{(1)})$ inequality (2.9), it is necessary to have all its neighboring boxes of sidelength $\ell$ non-singular. However, it may happen that one of the neighbors, $\uB_\ell(\uv^{(2)})$, is itself singular. In such a case, we can consider a bigger box $\uB_{2\ell-1}(\uv^{(1)}) \supset \uB_\ell(\uv^{(1)})$ as "insufficiently good" and try its neighbors, $\uB_\ell(\uv^{(3)})$, with $\dist [\uB_{2\ell-1}(\uv^{(3)}), \uB_\ell(v^{(2)}) ] = 1$; again, one of these boxes can be singular, in which case we obtain a finite sequence of singular boxes which we will call a singular chain:
$$
\uB_\ell(\uv^{(1)}), \; , \ldots, \uB_\ell(\uv^{(n)}), \; n\ge 1.
$$

\begin{figure}[h]
\center
    \includegraphics[height=0.4\textheight]{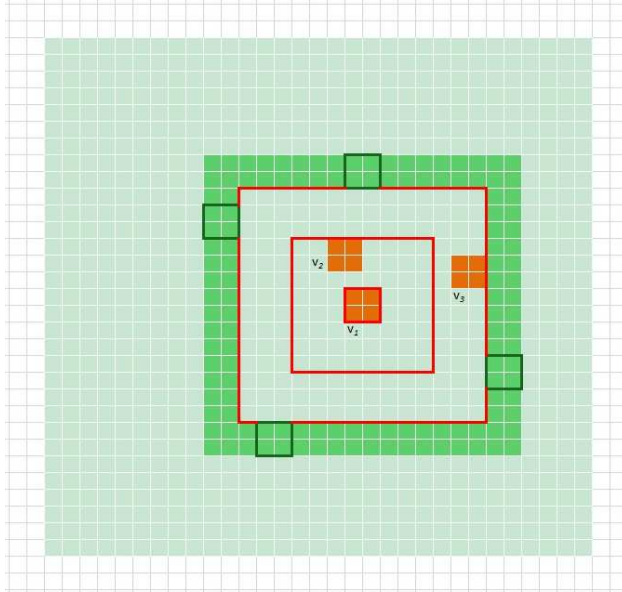}
    \caption{A singular cluster with 3 singular boxes (orange) centered at $\uv^{(1)}, \uv^{(2)}, \uv^{(3)}$. Neighboring boxes inside the green annular area are NS (four of these NS-neighbors are singled out with dark green border)\label{fig:fig01} }
\end{figure}
\pn
Observe that, by construction, any two boxes in the above singular chain are disjoint. Further, in some situations (e.g., in the multi-particle MSA scheme) one may need to have elements of a singular chain at a certain distance, e.g.,
$$
\dist [\uB_\ell(\uv^{(i)}), \uB_\ell(\uv^{(j)}) ] = b\ell, \; 1 \le i\neq j \le n.
$$
Starting with one singular box, we can construct a maximal singular chain. It is clear that if $\uB_L(u)$ contains no singular chain with $>n$ elements, $n\ge 1$, then for any point $\ux\in \uB_{L - 2n\ell}(\uu)$ (i.e., for any point not too close to the boundary of the box $\uB_L(\uu)$) admits the following inequality holds true:
$$
 \cR_{\uB_L(\uu)}(\uv,\uy;E) \le q \, \mymax{\uw: \|\uw - \uv\|=(A+1)\ell-1} \; \cR_{\uB_L(\uu)}(\uw,\uy;E),
$$
with $A = A(\uv) \le 2n$.

We will call a maximal singular chain a singular cluster.

It is worth mentioning that a box $\uB_L(\uu)$ may contain, in principle, several singular clusters, i.e., several maximal singular chains, and these clusters may contain different number of elements (disjoint singular boxes). For our purposes, it is not necessary to have singular clusters non-overlapping, although it is always possible, by making properly unions of singular boxes and surrounding such unions in larger boxes ("boxed singular clusters"), to construct a finite number of non-overlapping boxes such that
\psn
(i) no box of sidelength $\ell$ outside these boxed singular clusters  is singular;
\psn
(ii) any box of sidelength $\ell$ adjacent to the boundary of boxed singular clusters is non-singular;
\psn
(iii) if $\uB_L(\uu)$ does not contain any family of $\ge n$ non-overlapping singular boxes, then  the sum of diameters of all boxed singular clusters is bounded by $C(d) n\ell$, where $C(d)$ is a geometric constant.

Below we assume that singular clusters are constructed as described above, although such a construction is not unique. The most important property is (iii), which says that all singular boxes can be covered by a relatively small number of lattice cubes of size of order $O(n\ell)$, where $n$ is the maximal number  of possible families of disjoint singular boxes of size $\ell$.

\subsection{Subharmonicity of Green functions}

Given a box $\uB_L(\uu)$,  fix $E\in\R$ and define a function $f:\uB_L(\uu) \to \R_+$ by
$$
f(\ux) = \mymax{\uy\in \pt^- \uB_L(\uu)} \cR_{\uB_L(\uu)}(\ux,\uy;E).
\eqno(2.12)
$$
\pmn
Suppose that $\uB_L(\uu)$contains one or more singular clusters and define a set $\cS$ as the union of all singular clusters.  Then, by virtue of (2.10), for any lattice point $\ux\not\in\cS$ we have
$$
\cR_{\uB_L(\uu)}(\uu,\uy;E)| \le q \;\mymax{\uv:\, \|\uu - \uv\| = \ell-1} \; \cR_{\uB_L(\uu)}(\uv,\uy;E),
\eqno(2.13)
$$
\psn
while for points $\ux\in\cS$ we have, respectively,
$$
\cR_{\uB_L(\uu)}(\uu,\uy;E)| \le q \;\mymax{\uv:\, \ell \le \|\uu - \uv\| = 2\ell-1} \; \cR_{\uB_L(\uu)}(\uv,\uy;E),
\eqno(2.14)
$$
\psn
with the same value of $q$. Obviously, if $\cS=\varnothing$, then Eqn (2.13) can be used for all $\ell$-boxes inside $\uB_L(\uu)$, which only makes our estimates simpler.

In order to formalize such a property of a function $f$, we give the following
\pmn
\begin{Def}\label{Def_SH}
Consider a box $\uB_L(\uu)$ and a subset thereof $\cS\subset \uB_L(\uu)$. A function
$f:\, \uB_L(\uu)\to\R_+$ is called $(q,\ell,\cS)$-subharmonic if for all points $\ux\in\uB_L(\uu)\setminus\cS$ with
$\dist[\ux, \pt^-\uB_L(\uu) ] \ge \ell$ we have
$$
f(\ux) \le q \, \mymax{\uw:\, \|\uw-\ux\|=2\ell-1} f(\uw),
\eqno(2.15)
$$
\psn
and for every point $\ux\in\cS$ there exists an integer $\rho(\ux) \in[\ell, A\ell]$ and
$$
f(\ux) \le q \, \mymax{\uw:\, \rho(\ux) \le \|\uw-\ux\| \le \rho(\ux)+2\ell-1 } \;\;f(\uw).
\eqno(2.16)
$$
\end{Def}
\pmn
\textbf{Remark. } It is clear that, formally, we introduce the notion of $(\ell, q, \cS, A)$-subharmonicity.
The parameter $A$ is dropped for notational simplicity only, and this should not lead to any ambiguity.

We see that under the above assumptions upon the box $\uB_L(\uu)$,  the function
$$
f(\ux) := \mymax{\uy\in \pt^- \uB_L(\uu)} \cR_{\uB_L(\uu)}(\ux,\uy;E)
$$
\pmn
is $(q,\ell,\cS)$-subharmonic with $\cS $ defined as a union of all singular clusters and
$$
q = e^{-\gamma(m,\ell)} e^{\ell^\beta} C'(d) (n\ell)^{d-1}.
$$
Moreover, it is not difficult to see that if any family of disjoint singular boxes
$$
\uB_\ell(\uv^{(1)}), \uB_\ell(\uv^{(2)}), \ldots, \uB_\ell(\uv^{(j)}) \subset \uB_L(\uu)
$$
contains at most $n$ elements, i.e. $j\le n$, then the above function $f$ is $(q,\ell,\cS)$-subharmonic
with some set $\cS$ (which is \textbf{\textit{not}} defined in a unique way, in general) contained in a union
of annular areas
$$
\cA(\cS) := \bigcup_{i=1}^j \cA_i, \; \cA_i = \uB_{b_i}(u) \setminus \uB_{a_i}(u)
\eqno(2.17)
$$
with $0 < a_1 < b_1 < a_2 \ldots < a_j < b_j < L,$  $W(\cS) := \sum_{i=1}^j (b_i - a_i) \le 2n\ell$.
We will call $W(\cS)$ the (total) width of the singular area $\cA(\cS)$. If the annular covering $\cA(\cS)$ is chosen in a minimal way, then $W(\cS)$ is uniquely defined.

In the next subsection, we will establish a general bound for subharmonic functions, making abstraction of exact values of parameter  $q$.

\begin{figure}[h]
\center
    \includegraphics[height=0.4\textheight]{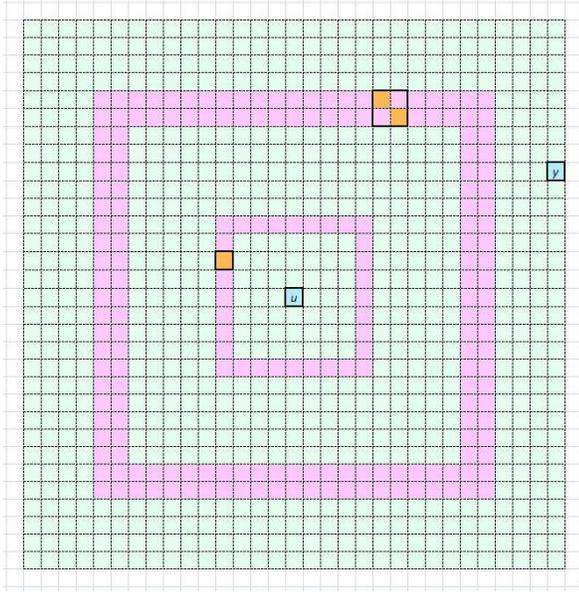}
    \caption{An example of a box $\uB_L(\uu)$ with two singular clusters (singular boxes are orange) covered by two annular areas (pink) \label{fig:fig02} }
\end{figure}

\subsection{Radial descent and decay of subharmonic functions}

The following elementary statement is an adaptation of Lemma 4.3 from \cite{C08}
\begin{Lem}\label{Lem_Radial}[Radial Descent Lemma]
Let $f$ be the function defined in (2.12) which is $(q,\ell,\cS)$-subharmonic in a box $\uB_L(\uu)$, covered by a union of annual areas $\cA(\cS)$, defined in (2.17), of total width $W(\cS)$. Then we have
$$
f(\uu) \le q^{(L - W(\cS)- 3\ell)/\ell} \cM(f, \uB_L(\uu)).
\eqno(2.18)
$$
\end{Lem}
The proof can be found in \cite{C08}; it is fairly straightforward.

\subsection{Application to the decay of Green functions}

It is readily seen that Lemma \ref{Lem_Radial} applied to the functions $f(\uu) = \cR(\uu,\uv;E)$ leads to the following
\begin{Lem}\label{Lem_RadialGF} Fix a non-negative integer $n<\infty$ and suppose that a box  $\uB_L(\uu)$ is $E$-non-resonant and that any maximal family of $b$-distant $(E,m)$-singular boxes contains at most $n$ elements. Then $\uB_L(\uu)$ is $(m,E)$-non-singular:
$$
\max_{\uy\in\pt^- \uB_L(\uu)}|G_{\uB_L(\uu)}(\uu,\uy;E)| \le \myexp{ - \gam(m,L)}.
$$
\end{Lem}

\pmn
\textbf{N.B.:} It is clear from our above analysis that all arguments, as well as the statement of Lemma
\ref{Lem_RadialGF}, remain valid for two-particle cubes in $\R^d$ (resp., two-particle boxes in $\Z^{2d}$. Indeed, apart from the difference in the value of the dimension and the additive structure of the potential
$\uV(x_1,x_2) = V(x_1) + V(x_2)$, the two-particle Hamiltonians similar form. Neither of these differences is crucial to our analysis, for the dimension can be arbitrary, and a particular structure of the potential is not used at all.

Note also that our analysis of $(\ell,q,\cS)$-subharmonic functions  is purely "deterministic" and does not rely upon any probabilistic assumption relative to the random external potential $V(x;\om)$.
\pmn
\textit{This concludes our reduction of the deterministic part of the continuous MSA to the lattice version thereof. The rest of the proof of exponential decay of Green functions is conducted in terms of the auxiliary
lattice model. The exponential decay of eigenfunctions is then deduced from that of
Green functions in a standard way. A reader familiar with \cite{CS09A} may notice that subsequent sections are straightforward adaptations of corresponding parts of \cite{CS09A}; they do not contain truly novel ideas or techniques, compared to \cite{CS09A}. }
\pmn
\pmn
\pmn

\section{Partial decoupling and tunneling in two-particle boxes}

Unlike the single-particle MSA, its two-particle counterpart proposed in \cite{CS09A} has to address the following difficulty of multi-particle models: the probabilistic dependence between the values of the potential
$\uV(\ux;\om) = V(x_1;\om) + V(x_2;\om)$ and $\uV(\uy;\om)=V(y_1;\om) + V(y_2;\om)$ does not decay with the distance $\|\ux - \uy\|$. However, a weaker form of "decoupling" in the potential $\uU(\ux)+\uV(\ux;\om)$ takes place for sufficiently distant points in the multi-particle configuration space. Such a decoupling, sufficient for the purposes of the two-particle MSA, makes use of the following elementary geometric statement (cf. \cite{CS09A}):
\pmn

\begin{Lem}\label{LemGeomTwoDist}

Let be $L > r_0$ and consider two interactive boxes, $\BLam_L(\uu')$ and
$\BLam_L(\uu'')$, with $dist(\BLam_L(\uu'), \BLam_L(\uu''))>8L$. Then
$$
\Pi \BLam_L(\uu'') \cap \Pi \BLam_L(\uu'') = \varnothing.
$$
\end{Lem}
The proof is straightforward and can be found in \cite{CS09A}.
\psn

Furthermore, in order to estimate the probability of simultaneous $(E,m)$-singularity of two $8L$-distant
cubes, we will making use of well-known results of the single-particle MSA, by introducing introduce the following

\begin{Def}\label{DefBad}
Given a bounded interval $I\subset \R$ and $m>0$, a single-particle box
$\Lam_{L_k}(u)$ is called $m$-tunneling ($m${\rm{-T}}, for short) if
$
\exists\, E\in I$  and disjoint boxes $\Lam_{L_{k-1}}(v_1)$,
$\Lam_{L_{k-1}}(v_2) \subset \Lam_{L_{k}}(u)$  which are $(E,m)${\rm-S}.
A two-particle box of the form
$\BLam_{L_k}(\uu) = \Lam_{L_{k-1}}(u_1) \times \Lam_{L_{k-1}}( u_2)$,
with $\uu = (u_1, u_2)$, is called   $m$-tunneling ($m${\rm{-T}})
if either $\Lam_{L_{k-1}}( u_1)$ or $\Lam_{L_{k-1}}( u_2)$ is $m$-tunneling. Otherwise, it is
called $m$-non-tunneling ($m${\rm{-NT}}, for short).
\end{Def}

It is worth mentioning that, while the notion of $m$-tunneling is, formally, defined for an arbitrary two-particle box, it is actually useful only in the case of a non-interactive box, where the spectral problem admits separation of variables,
and so is reduced to two single-particle spectral problems.

The following statement is a reformulation of well-known results of the single-particle MSA (cf. \cite{St01}
and bibliography therein), so its proof is omitted.

\begin{Lem}\label{LemTun}

Under the assumptions (E1--E4) upon the external (single-particle) external random potential $V(x;\om)$,
$$
\pr{ \Lam_{L_k}(u) \text{ is } m{\rm-T} }\le L_k^{-q'}
$$
where $q' = q'(\eta^*)$, $\eta^* := E_1^* - E_0^*>0$, can be chosen so that
$q'(\eta^*) \to +\infty$ as $\eta^* \downarrow 0$. Respectively, for a two-particle box
$\BLam_{L_k}(\uu) = \Lam_{L_{k-1}}(u_1) \times \Lam_{L_{k-1}}( u_2)$ we have
$$
\pr{ \BLam_{L_k}(u) \text{ is } m{\rm-T} }
\le  \sum_{j=1}^2 \pr{ \Lam_{L_k}(u_j) \text{ is } m{\rm-T} }  \le 2L_k^{-q'}.
$$
\end{Lem}

\section{ Reduction of the localization problem to the MSA }\label{SecReduct}
\pmn
\begin{Thm}\label{ThmReduct}
Suppose that for some $m>0$ and all $k\ge 0$ the following bound holds true: for any pair of $L_k$-distant two-particle boxes
$\uB_{L_k}(\uu')$ and $\uB_{L_k}(\uu'')$,
$$
\pr{\exists\, E\in[E_0^*, E_1^*]:\,  \text{$\uB_{L_k}(\uu')$ and $\uB_{L_k}(\uu'')$ are $(E,m)$-S} } \le L_k^{-2p}.
\eqno(\ref{SecReduct}.1)
$$
Then with probability one, the spectrum of operator $\uH(\om)$ in $[E_0^*, E_1^*]$ is pure point, and for any EF $\BPsi_j(\ux;\om)$ with $E_j(\om)\in [E_0^*, E_1^*]$, we have, for any $\uv\in\Z^{2d}$:
$$
\| \one_{\uC(\uv)} \, \BPsi_j(\cdot;\om)| \le C_j(\om) e^{-m\|\uv\|}.
\eqno(\ref{SecReduct}.2)
$$
\end{Thm}

For the reader's convenience, we give the proof of the above theorem in Section \ref{SecProofThmReduct}. All its ingredients can be found in \cite{CS09A} (as far as the two-particle structure of the Hamiltonian is concerned)) and in \cite{St01}.
\pmn

\textit{Therefore,  Anderson localization will be established, once we prove the main probabilistic bound of the MSA given by Eqn (\ref{SecReduct}.1)}.
\pmn

As usual in the MSA, the probabilistic bound (\ref{SecReduct}.1) is first established for $k=0$ initial length scale estimates), and then proved inductively for all $k\ge 1$.

The proof of   the initial length  scale estimate  is completely analogous to that in the conventional, single-particle
localization theory, and is omitted for this reason. Indeed, the reader may check that the arguments used, e.g.,
in \cite{St01} (cf. Ch. 3.3, pp. 90--98) do not use any assumption on the structure of the external potential
which is not satisfied in the two-particle (actually, even $N$-particle, with $N\ge 1$) model. The basis for these initial scale estimates is the well-known Combes-Thomas bound (cf. \cite{CT73}), combined with the fact that we consider energies $E\in[E_0^*,E_1^*]$ sufficiently close to the lower edge $E_0^*$ of the spectrum.

So, in the rest of the paper, we focus on the inductive proof of the bound (\ref{SecReduct}.1). To this end, we consider two kinds of boxes:
\psn
(i) \textit{non-interactive} boxes $\uB_L(\uu) = \BLam_L(\uu)\cap \Z^{2d}$ where the interaction potential vanishes: $\uU|_{\BLam_L(\uu)} \equiv 0$;
\pmn
(ii) \textit{interactive} boxes $\uB_L(\uu) = \BLam_L(\uu)\cap \Z^{2d}$ where the interaction potential is not identically zero on
$\BLam_L(\uu)$.
\pmn

This gives rise to three categories of \textit{pairs} of (sufficiently distant) boxes:
\pmn
\textbf{(I)} Two non-interactive boxes.
\pmn
\textbf{(II)} Two interactive boxes.
\pmn
\textbf{(III)} A mixed pair of one interactive  and one non-interactive box.

\pmn
These three cases will be treated separately in sections \ref{Sect_I}, \ref{Sect_II} and \ref{Sect_III}, respectively.
\psn

By virtue of Theorem \ref{ThmReduct},   Anderson localization (cf. Theorem \ref{ThmMain}) will be proven for the two-particle system in $\R^d$ with an alloy-type external random potential, verifying conditions (D), (E1)-(E4) given in Section \ref{intro},  once the bound (\ref{SecReduct}.1) is established in all cases (I)-(III).

\pmn
\textbf{Remark.} For the sake of notational simplicity, below we will call a box
$\uB_L(\uu) = \BLam_L(\uu)\cap\Z^{2d}$ $E$-non-resonant (resp., $E$-non-resonant) iff the corresponding cube
$\BLam_L(\uu)\subset\R^{2d}$ is $E$-resonant (resp., $E$-non-resonant).

\section{Pairs of non-interactive boxes}\label{Sect_I}

We begin with an auxiliary result  about non-interactive boxes, which was earlier used in \cite{CS09A}, \cite{CS09B}.
For  the reader's convenience, we give its proof (which is straightforward) in the Appendix.

\begin{Lem}\label{LemNITRoNS}
Suppose that a two-particle box $\uB_{L_{k+1}}(\uu)$ is $E$-non-resonant and satisfies the following property: for any pair of
sub-boxes $\uB_{L_k}(\uv'), \uB_{L_k}(\uv')\subset \uB_{L_{k+1}}(\uu)$ with
$\dist [\uB_{L_k}(\uv'), \uB_{L_k}(\uv'') ] > 8L_k$, either $\uB_{L_k}(\uv')$ or $\uB_{L_k}(\uv')$ is $(E,m)$-non-singular. Then $\uB_{L_{k+1}}(\uu)$ is also $(E,m)$-non-singular.
\end{Lem}

\textit{Proof of (\ref{SecReduct}.1) for a pair of non-interactive boxes}.

Consider a pair of two-particle non-interactive boxes $\uB' =\uB_{L_{k+1}}(\uu')$, $\uB'' =\uB_{L_{k+1}}(\uu'')$,
and introduce the events
$$
\begin{array}{l}
\uT= \myset{\text{either $\Lam'$ or $\Lam''$ is $m$-T } }, \\
\uR= \myset{\exists\, E\in [E_0, E_1]:\; \text{both $\Lam'$ and $\Lam''$ are $E$-R } },\\
\uS= \myset{\exists\, E\in [E_0, E_1]:\; \text{both $\Lam'$ and $\Lam''$ are $(E,m)$-S } }.
\end{array}
$$
Then we can write
$$
\pr{\uS} \le \pr{\uT} + \pr{ \uS \cap \uT^c}.
$$
Owing to Lemma \ref{LemTun}, we have
$$
\pr{\uT} \le \pr{ \text{$\Lam'$  is $m$-T } } + \pr{\text{$\Lam''$ is $m$-T } } \le 2 \cdot 2 L_k^{-q'},
$$
where $q'>0$ can be chosen arbitrarily large, provided that $E_1^*-E_0^*$  is sufficiently small. So, we can pick $q' \ge q$ with $q>0$ given in the Wegner-type bound (W2).  Further, by two-volume Wegner-type estimate (W2), we have
$$
\pr{ \uR} < L_k^{-q}.
$$
By virtue of the NITRoNS (Lemma \ref{LemNITRoNS}), $\uS \cap \uT^c \subset \uR$,  Now,  using (W2), we obtain
$$
\pr{\uS} \le 4 L_k^{-q'} + \pr{\uR} \le L_k^{-q'} + L_k^{-q} \le 2 L_k^{-q}
 < L_k^{-2p},
$$
owing to our choice of parameter $q (> 3p+9)$, for all sufficiently large $L_k$. Thus, the bound (\ref{SecReduct}.1) is proven
for distant pairs of non-interactive boxes.

\section{Pairs of interactive boxes}\label{Sect_II}

Consider again the following events:
$$
\begin{array}{l}
\uR= \myset{\exists\, E\in [E_0, E_1]:\; \text{both $\uB'$ and $\uB''$ are $E$-R } },\\
\uS= \myset{\exists\, E\in [E_0, E_1]:\; \text{both $\uB'$ and $\uB''$ are $(E,m)$-S } }.
\end{array}
$$
Using the Wegner-type bound (W2) and our condition $q>3p+9$ , we see that
$$
\pr{ \uS } \le \pr{\uR} + \pr{ \uS \cap \uR^c} \le \half L_k^{-2p} + \pr{ \uS \cap \uR^c}.
\eqno(6.1)
$$
Within the event $\uR^c$, either $\uB'$ or $\uB''$ is $E$-non-resonant. Without loss of generality, assume that
$\uB'$ is $E$-non-resonant.

By virtue of the Radial Descent Lemma, if $\uB'$ is $(E,m)$-singular, but $E$-non-resonant, then it must contain
a singular cluster of $2M+1 \ge 5$ (with $M=2$) distant sub-boxes $\uB_{L_k}(\uu_j)$, $j=1, \ldots, 2M+1$.

Consider the following events:
$$
\begin{array}{l}
\uS'_I= \myset{\text{$\uB'$  contains at least two  $(E,m)$-S non-interactive boxes} }, \\
\uS'_{NI}= \myset{\text{$\uB'$  contains at least $2M \ge 4$  $(E,m)$-S interactive boxes} }.
\end{array}
$$
Obviously, $\uS' \subset \uS'_I \cup \uS'_{NI}$.

Reasoning as in Section \ref{Sect_I}, we conclude that $\pr{\uS'_I} \le 2L_k^{-q}$.

Further, suppose that  $\uB'$  contains at least $2M$  $(E,m)$-singular distant interactive boxes
$\uB_{L_k}(\uu_j)$, $j=1, \ldots, 2m$. Owing to Lemma \ref{LemGeomTwoDist}, the external potential
samples in boxes $\uB_{L_k}(\uu_j)$ are independent. The situation here is completely analogous to that in the single-particle theory, and we can write that
$$
\begin{array}{l}
\pr{\uS'_{NI}} \\
\le L_{k+1}^{2M(d+\alpha^{-1})} \prod_{i=1}^M \pr{\exists\, E\in[E_0,E_1]:\; \text{$\uB_{L_k}(\uu_j)$ and $\uB_{L_k}(\uu_j)$ are  $(E,m)$-S} } \\
\le L_{k+1}^{2M(d+\alpha^{-1})} \left( L_k^{-2p}\right)^M < \half L_{k+1}^{-2p},
\end{array}
$$
as long as $p > \frac{3d}{2} + 1$, with $M=2$, and $L_0$ (hence, every $L_k$, $k\ge 1$) is sufficiently large.
Taking into account Eqn (\ref{Sect_II}.1), we see that
$$
\pr{\uS} \le \half L_{k+1}^{-2p} + \half L_{k+1}^{-2p} = L_{k+1}^{-2p},
$$
yielding the bound (\ref{SecReduct}.1) for pairs of (distant) interactive boxes.

\section{Mixed pairs of boxes}\label{Sect_III}

It remains to derive the bound (\ref{SecReduct}.1) in case (III), i.e., for mixed pairs
of two-particle boxes: an interactive box $\uB_{L_{k+1}}(\ux)$ and a non-interactive box $\uB_{L_{k+1}}(\uy)$.
Here we use several properties which have
been established earlier in this paper for all scale lengths,
namely, \textbf{(W1)}, \textbf{(W2)}, \textbf{NITRoNS}, and the bound (\ref{SecReduct}.1)
for pairs of (distant) non-interactive boxes, in Section \ref{Sect_I}.

Consider the following  events:
$$
\begin{array}{l}
\uS = \Big\{ \exists \, E\in I:\,{\rm{both}}\;
\uB_{L_{k+1}}(\ux),\;\;\uB_{L_{k+1}}(\uy)\;
\text{ are } (E,m)\text{-S}\Big\}\,, \\
\uT = \Big\{\hbox{  $\uB_{L_{k+1}}(\uy)$ is $m_{0}$-T} \Big\},\\
\uR = \Big\{\exists \, E\in I:\, \text{ neither }
\uB_{L_{k+1}}(\ux) \text{ nor }
\uB_{L_{k+1}}(\uy)\text{ is } (E,J){\text{-NR}} \Big\}.
\end{array}
$$
As before, we have
$$
\pr{ \uT } \le L_{k+1}^{-q'} \le L_{k+1}^{-q},
\quad
\pr{\uR} \leq L_{k+1}^{-q}.
\eqno (7.1)
$$
Further,
$$\pr{\uS} \leq \pr{\uT} + \pr{\uS\cap \uT^{\rm c}} \leq
\frac{1}{4} L_{k+1}^{-2p} + \pr{\uS\cap \uT^{\rm c}},
$$
and for the last term in the RHS we have
$$
\pr{\uS \cap \uT^{\rm c}}
\leq \pr{\uR} + \pr{\uS\cap \uT^{\rm c}\cap \uR^{\rm c}}
\leq L_{k+1}^{-q+2}
+ \pr{\uS\cap \uT^{\rm c}\cap \uR^{\rm c}}.
$$
Within the event
$\uS\cap \uT^{\rm c} \cap \uR^{\rm c}$, either $\uB_{L_{k+1}}(\ux)$ or $\uB_{L_{k+1}}(\uy)$
is $E$-non-resonant. It must be the interactive box $\uB_{L_{k+1}}(\ux)$. Indeed, by
\textbf{NITRoNS}, had box $\uB_{L_{k+1}}(\uy)$ been both $E$-non-resonant and $m$-non-tunneling,
it would have been $(E,m)$-non-singular, which is not allowed within the event $\uS$.
So, the box $\uB_{L_{k+1}}(\ux)$ must be $E$-non-resonant, but $(E,m)$-singular:
$$
\uS\cap \uT^{\rm c} \cap \uR^{\rm c} \subset \{\exists \, E\in I:\;\uB_{L_{k+1}}(\ux)
\text{ is } (E,m){\text{-S}} \text{ and } E{\text{-NR}} \}.
$$
However, applying the Radial Descent Lemma, we see that
$$
\begin{array}{r}
\{\exists \, E\in I:\,
\uB_{L_{k+1}}(\ux) \text{ is }(E,m){\text{-S}} \text{ and }
E{\text{-NR}}\}\qquad\qquad\\
\subset \{\exists \, E\in I:\, K(\uB_{L_{k+1}}(\ux);E) \geq J+1\}.
\end{array}
$$
Therefore,
$$
\begin{array}{cl}
\pr{\uS\cap \uT^{\rm c}\cap \uR^{\rm c}}
&\leq \pr {\exists \, E\in I:\, K(\uB_{L_{k+1}}(\ux)
;E) \geq 2\ell+2 }\\
\;&\leq 2L_{k+1}^{-1}\,L_{k+1}^{-2p}.\end{array}
$$
Finally, we get, with $q'':=q/\alpha = 2q/3 > 2p+6$,
$$\begin{array}{cl}
\pr{\uS} &\leq \pr{\uS\cap \uT} + \pr{\uR}  + \pr{\uS\cap \uT^{\rm c}
\cap \uR^{\rm c}}\\
\;&\leq \half L_{k+1}^{-2p} + L_{k+1}^{-2p-2} + 2L_{k+1}^{-1} \, L_{k+1}^{-2p}
\leq L_{k+1}^{-2p},\end{array}
$$
This completes the proof of bound (\ref{SecReduct}.1).

\pmn

\textit{Therefore, Theorem \ref{ThmMain} is also proven and the Anderson localization established for a two-particle model satisfying hypotheses \textbf{(D)} and \textbf{(E1)} -- \textbf{(E4)}}.

\section{Appendix. Proof of NITRoNS principle }\label{SecNITRoNS}

 Here we give the proof of Lemma \ref{LemNITRoNS}. Recall  that we consider  operator $\uH^{\BLam_{L_k}(\uu)}$ in a cube $\BLam_{L_k}(\uu)$ and "single-particle" operators $H^{\Lam^{L_k}(u')}$, $H^{\Lam^{L_k}(u'')}$.
 Let $\{\ffi_a,\lam_a\}$ be normalized eigenfunctions  and the respective eigenvalues of $H^{\Lam^{L_k}(u')}$. Similarly,  let $\{\psi_b,\mu_b\}$ be normalized eigenfunctions  and the respective eigenvalues of $H^{\Lam^{L_k}(u'')}$.

 Consider the Green functions
 $\uG(\uv,\uy;E_j) \equiv \uG^{\BLam_{L_k}(\uu)}(\uv,\uy;E_j)$, $\uv,\uy\in \uB_{L_k}(\uu)$.
Observe that, since the external potential is non-negative, so are the eigenvalues $\{\lam_a\}$ and $\{\mu_b\}$. Therefore, if $E\le E^*_1$, then we also have
$E - \lam_a \le E^*_1$, $E - \mu_b \le E^*_1$, for all $\lam_a$ and $\mu_b$.

By the hypothesis of the lemma, $\BLam_{L_k}(\uu)$ is $E$-non-resonant. Therefore, for all $\lam_a$, the $1$-particle box $\Lam_{L_k}(u'')$ is
$(E-\lam_a)$-non-resonant. By the assumption of $m$-non-tunneling, $\forall\, E\in [E^*_0, E^*_1]$ box
$\Lam^{L_k}(u'')$ must not contain two disjoint $(E-\lam_a, m)$-singular sub-boxes of size $L_{k-1}$.
Therefore, the Radial Descent Lemma implies that $\Lam_{L_k}(u'')$ is $(E-\lam_a)$-non-singular, yielding the required upper bound.

Let us now prove the second assertion of the lemma.  If $\uu=(u',u'')$ and $\uv=(v',v'')\in\pt \BLam_{L_k}(\uu)$, then either
$\|u' - v'\|=L_k$, or $\|u'' - v'' \|=L_k$. In the former case we can write
$$
\begin{array}{cl}
 \uG(\uu, \uv;E) &
= \displaystyle \sum_{a} \ffi_a(u') \ffi_a(v') \, \sum_{b} \, \frac{ \psi_b(u'') \psi_b(v'') }{ (E - \lam_a) - \mu_b } \\ \\
\;&= \displaystyle\sum_{a} \ffi_a(u') \ffi_a(v') \,  \,  G^{\Lam_{L_k}(u'')}(u'', v''; E - \lam_a).
\end{array}
\eqno(8.1)
$$
As mentioned above, $E - \lam_a \le E_1^*$. In fact, by Weyl's law, $E - \lam_a \to -\infty$ as $a\to\infty$.
More precisely, for all $a \ge a^* =  C^*|\Lam_{L_k}(u')|$ (with constant $C^*$ given by the Weyl's law),  we have
$E-\lam_a \le -m^*$, where $m^*>0$ can be chosen arbitrarily large,  and, therefore, $E-\lam_a<0$ is far away from the (positive) spectrum:
$$
\dist [\Sigma(H^{\Lam_{L_k}(u')}), E-\lam_a] = | E_0 -(E-\lam_a) |  \ge m^*.
$$
By virtue of the Combes--Thomas estimate, if $E - \lam_a \le -m^*$ and  $m^*>0$ is large enough, then
$$
\max_{v'\in\pt \Lam_{L_k}(u')}
\, \| \one_{C(u'')} G^{\Lam_{L_k}(u'')}(E - \lam_a) \one_{C(v'')}\| \le e^{-m\|u' - u''\|} \le e^{-m L_k}.
$$
On the other hand, given any non-negative number $m^*$, one can consider from the beginning the energy interval $[-m^*,E_1^*]$ instead of $[E_0^*,E_1^*]$. Considering negative energies is fictitious, yet the standard, single-particle MSA would imply, formally, all required probabilistic MSA estimates for such a larger interval $[-m^*,E_1^*]$. The same is true, of course, for the two-particle MSA.


Therefore, an infinite sum over $a$ in (8.1) can be divided into two sums:
$$
\begin{array}{cl}
 \uG(\uu, \uv;E) &= \displaystyle\left( \sum_{a \le a^*} +  \sum_{a > a^*} \right) \ffi_a(u') \ffi_a(v') \,  \,  G^{\Lam_{L_k}(u'')}(u'', v''; E - \lam_a),
\end{array}
\eqno(8.1')
$$
where the (infinite) sum $ \sum_{a > a^*} (\cdot)$ can be made smaller than, for example, $e^{-2mL_k}$, by choosing
$a^*$ large enough, thus making  $m^*>0$ large enough. On the other hand, the first sum,  $ \sum_{a \le a^*} (\cdot)$, contains a finite number of terms: $O(L_k^d)$.

Since $\|\ffi_a \|=1$ for all $a$, we see that
$$
\begin{array}{ll}
\| \one_{\uC(\uu)} \uG(E) \one_{\uC(\uv)}\| \\
 \leq \,e^{-2mL_k} +  C'\left| \Lam_{L_k}(u') \right| \,
\mymax{a \le a^*} \| \one_{C(u'')} G^{\Lam_{L_k}(u'')}(E - \lam_a) \one_{C(v'')}\|  \\ \\
  \le \, C'' \, (2L_k)^d \, e^{-mL_k},
\end{array}
\eqno(8.2)
$$
owing to the $(E-\lam_a,m)$-non-singularity of the cube $\Lam_{L_k}(u'')$.

In the case where $\|u'' - v'' \|=L_k$, we can use the representation
$$
 \uG(\uu, \uv;E)
= \sum_{b} \psi_b(u'') \psi_b(v'') \,  \,  G^{\Lam_{L_k}(u')}(u', v'; E - \mu_b).
\eqno(8.3)
$$
\qed

\section{Appendix B. Proof of the Theorem \ref{ThmReduct}}\label{SecProofThmReduct}

\begin{Lem}\label{Lem331}
Fix an interval $I = [E_0, E_1]\subset \R$ and a sequence of positive numbers $\{L_k = (L_0)^{\alpha^k}\}$,
$L_0 >0$, $\alpha\in(1,2)$. Suppose that the bounds (4.1) are satisfied for all $k \ge 0$.

Then there exists a positive number $m$ and a subset $\Om_0\subset\Om$ with $\pr{\Om_0}=1$ such that for every $E\in I$ and $\om\in\Om_0$ and for every polynomially bounded function $\uf\in L^2_{loc}(\R^{2d})$
satisfying
$$
\| \one_{\uC(\uv}) \uf\|
\le C(\uf) \cdot \| \chiout_{\ell,\uv}\, \cR_{\ell,\uv}(E;\om) \, \chiint_{\ell,\uv}  \| \cdot
\|  \chiout_{\ell,\uv}\, \uf \|
\eqno(9.1)
$$
there exists $C = C(\uf, \om, m)$ such that
$$
\| \one_{\uC(\uv}) \uf\| \le C \, e^{- m \|\uv\|}
\eqno(9.2)
$$
and, more precisely,
$$
\limsup_{\|\uv\|\to\infty} \; \frac{\ln (\| \one_{\uC(\uv)} \uf\| )}{ \|\uv\|} \le -m.
\eqno(9.3)
$$
\end{Lem}

\proof

Let $R:\R^{2d}\to\R_+$ be the function given by $R(\uu) = \|\uu - S(\uu)\|$, where $S(u_1,u_2) = (u_2,u_1)$,
$\uu = (u_1,u_2)\in\R^{2d}$. Next, for every $k\in\N$, set
$$
b_k(\uu) = 1 + R(\uu) L_k^{-1}, \; \uM_k(\uu) = \BLam_{L_k}(\uu) \cup  S \left(\BLam_{L_k}(\uu) \right).
$$
Observe that for any $\uu\in\R^{2d}$ we have
$$
\forall  k\ge 0:\; \uM_k(\uu) \subset  \BLam_{b_{k+1}L_k}(\uu),
\qquad
\text{ and } \lim_{k\to\infty} b_k(\uu) = 1.
\eqno(9.4)
$$

Next, introduce annular subsets of the lattice
$$
A_{k+1}(\ux_0) = \BLam_{2b L_{k+1}}(\ux_0)\setminus \BLam_{2 L_{k}}(\ux_0) \cap \Z^{2d}
$$
centered at points $\ux_0\in\Z^{2d}\subset\R^{2d}$. Next, consider events
$$
\begin{array}{ll}
\uS_k(\uu) = & \{\exists\,E\in I, \ux\in A_{k+1}(\ux_0):\,   \BLam_{ L_{k}}(\uu) \text{  and } \BLam_{ L_{k}}(\ux)
       \text{ are }  (E,m)-S \}.
\end{array}
$$
Observe that, owing to the definition of $\uM_k(\uu)$, if $\ux\in \uA_{k+1}(\uu)$, then
$$
\dist[ \BLam_k(\uu),  \uM_k(\uu)] > 8L_k,
\eqno(9.5)
$$
and, by the hypothesis of the lemma,
$$
\pr{\uS_k(\uu) } \le \frac{  (2b_{k+1} L_{k+1} +1)^2d }{ L_k^{2p}}
\le  \frac{  (2b_{k+1} +1)^2d }{ L_k^{2p-2\alpha} }.
\eqno(9.6)
$$
Since $p>\alpha$, and by virtue of (9.4), $\sum_{k\ge 0} \pr{\uS_k(\uu) } < \infty$, and the event
$$
\uS_\infty(\uu) = \myset{ \uS_k(\uu) \text{ occurs infinitely many times } }
$$
has probability zero, by virtue of the Borel--Cantelli lemma. As a consequence, the event
$$
\uS_\infty = \bigcup_{\uu\in\Z^{2d} } \uS_\infty(\uu)
$$
also has probability zero, so that its complement
$$
\Om_0 =
\{\forall\,\uv\in\Z^{2d} \;\exists\, k_\uv(\om)\in\N\; \text{ such that } \forall \, k\ge k_\uv(\om) \;
\uS_{k}(\uv) \not\ni \om \}.
$$
has probability $1$.

The rest of the proof is purely "deterministic". Let $E\in I$, $\om\in\Om_0$ and $\uf\in L^2_{loc}(\R^{2d})$ a polynomially bounded function satisfying Eqn. (9.1). If $\uf \not = 0$, then there exists a lattice point $\ux_0$ such that $\|\one_{\uC(\ux_0)} \uf\|>0$; we pick such a point $\ux_0$ and fix it for the rest of the proof. The box $\BLam_{L_k}(\ux_0)$ cannot be $(E,m)$-nonsingular for infinitely many values of $k$, since it would imply that
$$
\|\one_{\uC(\ux_0)} \uf\| \le Const \, L_k^{2d-1} e^{-mL_k} \tto{k\to\infty}{0}{},
\eqno(9.7)
$$
hence, $\|\one_{\uC(\ux_0)} \uf\|=0$, in contradiction with our hypothesis. Thus, there exists some $k_0$ such that
$\forall\, k\ge k_0$ the cube $\BLam_{L_k}(\ux_0)$ cannot be $(E,m)$-singular. In turn, this means, by construction of the event $A_{k+1}$, that for any point $\ux\in A_{k+1}(\ux_0)$ the cube $\BLam_{L_k}(\ux_0)$ is $(E,m)$-nonsingular.

Further, set
$$
\uB_{k+1}(\ux_0) = \BLam_{\frac{2b}{1+\rho} L_{k+1}}(\ux_0)\setminus \BLam_{\frac{2}{1-\rho} L_{k}}(\ux_0)
\subset A_{k+1}(\ux_0).
$$
It is readily seen that for any $\ux\in \uB_{k+1}$, we have $\dist[x, \uB_{k+1}(\ux_0)] \ge \rho \, \|\ux - \ux_0\|$. Furthermore, if $\|\ux - \uu\| \ge L_0/(1-\rho)$, then $\exists\, k\ge 0$ such that $\ux\in \uB_{k+1}(\uu)$.

Now we see that for sufficiently big $k\ge 0$, the box $\BLam_{L_k}(\uu)$ is $(E,m)$-non-singular, so that
$E \not\in {\rm spec}\, (\uH^{\BLam_{L_k}(\uu)})$. Therefore, we can apply the GRI and obtain
$$
\|\one_{C_1(\ux) }\uf\| \le C(d) L_k^{2d-1} e^{-mL_k}
\max_{ \uv ... } \,  \|\one_{C_1(\uv) }\uf\|
\eqno(9.8)
$$

Pick a value $\trho\in(0,1)$ and write it as a product of the form $\trho =\rho \rho'$ with some
$\rho, \rho'\in(0,1)$. Pick also a number $b > 8 + 1 + \rho/(1-\rho)$. The above inequality (9.8) can iterated
at least $n_k:=((L_k+1)^{-1}\rho\|\ux - \uu\|$ times, producing the following bound:
$$
\|\one_{C_1(\ux)} \uf\| \le
\left(  C(d) L_k^{2d-1} e^{-mL_k}  \right)^{n_k} Const \, (1 + \|\uu\| + bL_{k+1})^t.
$$
Therefore, for $k$ big enough, if $\|\ux - \uu\| \ge L_k/(1-\rho)$, then
$$
\|\one_{C_1(\ux) }\uf \| \le  e^{ -\rho \rho' m \|\ux - \uu\|},
$$
so that
$$
\limsup_{\|\ux\|\to\infty} \frac{ \ln \|\one_{C_1(\ux)} \uf\| }{\|\ux\|} \le - \rho \rho' m.
$$
This completes the proof of Lemma \ref{Lem331}. \qed
\pmn
In the following statement, we treat individual realizations of the random Hamiltonian $\uH(\om)$. This possible owing to our assumption of boundedness of the random amplitude of "impurities", $V(x;\om), x\in\Z^d$. In a more general case, a similar statement can be proved with probability one with respect to the ensemble of potentials $V(x;\om)$. In fact, Lemma {Lem332} follows from a much more general statement from \cite{St01}, so we omit here its proof.

\begin{Lem}\label{Lem332}[Cf. Lemma 3.3.2 in \cite{St01}, Section 3.3]
Assume that $\uH(\om)$ satisfies hypotheses (D) and (E1)--(E4). Then the following properties hold true:
\psn
(A) For spectrally almost every $E\in \BSig(\uH(\om))$ there exists a polynomially bounded eigenfunction corresponding to $E$.
\psn
(B) For every bounded set $I_0 \subset \R$ there exists a constant $C = C(M,I_0)$ such that for every generalized eigenfunction $\BPsi$ of $\uH(\om)$ corresponding to $E\in I_0$ satisfies
$$
\| \one_{\uC(\uu)} \, \BPsi\| \le C \,
\| \one_{\uC(\uw)} \, (\uH^{\BLam}(\om) - E)^{-1} \one_{\uC(\uu)}\| \; \| \one_{\uC(\uw)} \, \BPsi\|
$$
where $\uH^{\Lam}$ is the restriction of  $\uH(\om)$ to $L^2(\BLam(\uu)$ with Dirichlet boundary conditions.
\end{Lem}

Now we are prepared to prove Theorem \ref{ThmReduct}. Indeed, by Lemma \ref{Lem332}, there is a set $\cE_0\subset I = [0, E_0^*]$ with the following properties:

\begin{itemize}
\item
$\forall\, E\in\cE_0$ there is a polynomially bounded eigenfunction $\BPsi$ of $\uH(\om)$ corresponding to $E$;

\item $I\setminus \cE_0$ is a set of measure zero for the spectral resolution of operator $\uH(\om)$.
\end{itemize}

Further, by Lemma \ref{Lem331}, every polynomially bounded generalized eigenfunction $\BPsi$ corresponding to
$E\in I$ is exponentially decaying, in the $L^2$-sense, and in particular, $\BPsi \in L^2(\R^{2d})$. This means that $E$ is actually an eigenvalue. Moreover, since the Hilbert space $L^2(\R^{2d})$ is  separable, this implies that the spectrum of $\uH(\om)$ is pure point and, as was just mentioned, all corresponding eigenfunctions   decay exponentially in the $L^2$-sense, as stated in the Theorem \ref{ThmReduct}. This concludes the proof. \qed

\pmn

\pmn{\bf  Acknowledgments.}  The authors  thank the Isaac Newton Institute and the organizers of the program "\textit{Mathematics and Physics of Anderson localization: 50 years after}"  for their hospitality. P.S. thanks the DFG (German Science Foundation) for financial support of   visits to Paris and Cambridge.

\bibliographystyle{amsalpha}

\begin{thebibliography}{A}

\bibitem[AGKW08]{AGKW08} M. Aizenman, F. Germinet, A. Klein, S. Warzel, \textit{On Bernoulli decompositions for random
variables, concentration bounds, and spectral localization.} - Probab. Theory Relat. Fields,
\textbf{143}, 219-–238 (2009).

\bibitem [AW08] {AW08} M. Aizenman, S. Warzel, \textit{Localization Bounds for Multiparticle
Systems.} - Preprint,  arXiv math-ph/0809:3436 (2008); to appear in:  Comm. Math. Phys.

\bibitem [BCH97] {BCH97}
  {J.M.}{Barbaroux}, {J.M.}{Combes} and  {P.D.}{Hislop}:
  Localization near band edges for random Schr\"odinger operators.
  {\it Helv. Phys. Acta}, {\bf 70}, 16--43 (1997).

\bibitem[BK05]{BK05}  J. Bourgain and C. Kenig: On localization in the continuous Anderson-Bernoulli model in higher dimension. {\it Inv. Math.}, {\bf 161}(2), 389--426 (2005).

\bibitem[BCSS08] {BCSS08} A. Boutet de Monvel, V. Chulaevsky, P. Stollmann, Y. Suhov:
\textit{Wegner-type bounds for a two-particle continuous Anderson model with an alloy-type external potential.}
- arXiv:math-ph/0821:2621 (2008).

\bibitem[BCS08] {BCS08} A. Boutet de Monvel, V. Chulaevsky, Y. Suhov:
\textit{Wegner-type bounds for a two-particle
Anderson model in a continuous space.} - arXiv:math-ph/0821:2627 (2008).

\bibitem [C08] {C08}  V. Chulaevsky: \textit{Localization with Less larmes: Simply MSA.}
- Preprint, arXiv:math-ph/0812.2634, 2008.

\bibitem [CS08] {CS08}  V. Chulaevsky, Y. Suhov:  \textit{Wegner bounds for a two-particle
tight binding model.} - Commun. Math. Phys., \textbf{283}, 479--489 (2008).

\bibitem [CS09A] {CS09A}  V. Chulaevsky, Y. Suhov: \textit{Eigenfunctions in a two-particle
Anderson tight binding model}. - Comm. Math. Phys., \textbf{289}, 701--723 (2009); arXiv math-ph/0810:2190 (2008).

\bibitem [CS09B] {CS09B}  V. Chulaevsky, Y. Suhov: \textit{Multi-Particle Anderson Localisation: Induction on the Number of Particles.} - Math. Phys. Anal. Geom., \textbf{12}, 117--139 (2009);  arXiv math-ph/0811.2530 (2008).

\bibitem[CH94]{CH94} {J.M.}{Combes} and  {P.D.}{Hislop}: Localization for some
  continuous, random Hamiltonians in d-dimensions, {\it J. Funct.
    Anal.}, {\bf 124}, 149--180 (1994).

\bibitem[CT73]{CT73} {J.M.}{Combes} and  {L.}{Thomas}: \textit{Asymptotic Behaviour of Eigenfunctions
for Multiparticle Schrodinger Operators}, Comm. Math. Phys., \textbf{34}, 251--270 (1973).

\bibitem[DS01]{DS01} {D.}{Damanik} and {P.}{Stollmann}: \textit{Multi-scale analysis
  implies strong dynamical localization}, GAFA, {\it Geom. funct. anal.}
  {\bf 11}, 11--29 (2001).

\bibitem [D87] {D87} H. von Dreifus, \textit{On the effect of randomness inferromagnetic models and Schr\"{o}dinger
operators.} PhD thesis, New York University, 1987.


\bibitem [DK89] {DK89}  H. von Dreifus, A. Klein: \textit{A new proof of Localization in the Anderson
Tight Binding Model}. - Commun. Math. Phys., \textbf{124}, 285--299 (1989).

\bibitem[FS83] {FS83} J. Fr\"{o}hlich, T. Spencer: \textit{Absence of diffusion in the Anderson tight
binding model for large disorder or low energy}. - Commun. Math. Phys., \textbf{88}, 151--184 (1983).

\bibitem [FMSS85]{FMSS85} J. Fr\"{o}hlich, F. Martinelli, E. Scoppola, T. Spencer: \textit{A constructive
proof of localization in Anderson tight binding model}. - Comm. Math. Phys., \textbf{101}, 21--46 (1985).


\bibitem[GK01]{GK01} F. Germinet, H. Klein: Bootstrap Multiscale   Analysis and Localization in Random Media. {\it Commun. Math. Phys.}  {\bf 222}(2), 	415-448 (2001).

 \bibitem[GMP77]{GMP77} {I.Ya.}{Goldsheid}, {S.A.}{Molchanov},  {L.A.}{Pastur}:
  Typical one-dimensional Schr\"{o}dinger operator has pure point
  spectrum, {\it Funktsional. Anal. i Prilozhen.} {\bf 11} (1), 1--10
(1977); Engl.  transl. in {\it Functional Anal. Appl.} {\bf 11}   (1977).

\bibitem[HM84]{HM84} {H.}{Holden} and  {F.}{Martinelli}: On the absence of
  diffusion for a Schr\"odinger operator on $L^2(R^{\nu})$ with a
  random potential. {\it Commun. Math. Phys.}  {\bf 93}, 197--217   (1984).

 \bibitem[KSS98A]{KSS98A} {W.}{Kirsch}, {P.}{Stollmann} and  {G.}{Stolz}:
  Localization for   random perturbations or periodic Schr\"odinger operators. {\it
    Random Operators and Stochastic Equations}, {\bf 8}(2), 153--180   (1998).

\bibitem[KSS98B]{KSS98B} {W.}{Kirsch}, {P.}{Stollmann }and  {G.}{Stolz}: Anderson
  Localization for random Schr\"odinger operators with Long Range
  Interactions. {\it Commun. Math. Phys.} {\bf 195}, 495--507 (1998).

\bibitem[K95]{K95} {F.}{Klopp}: Localization For Some Continuous Random
  Schr\"odinger operators. {\it Commun. Math. Phys.} {\bf 167}, 553--569 (1995).

\bibitem[KZ03]{KZ03} {F.}{Klopp}, {H.} {Zenk}: The integrated density of states for an interacting
multielectron homogeneous model. Preprint, Universit\'{e} Paris-Nord (2003).

\bibitem[KZ09]{KZ09} {F.}{Klopp}, {H.} {Zenk}:The Integrated Density of States for
an Interacting Multiparticle Homogeneous Model and Applications to the Anderson Model.
Adv. Math. Phys., \textbf{2009}, 1--15 (2009).



\bibitem[S88] {S88} T. Spencer: \textit{Localization for Random and Quasiperiodic Operators}. - J.
Statist. Phys., \textbf{ 51}, No. 5/6,  1009--1019 (1988).

\bibitem [St01] {St01} P. Stollmann, \textit{ Caught by disorder}. - Birkh\"{a}user, 2001.



\end{thebibliography}

\end{document}